\documentclass{egpubl}
\usepackage{pg2025}

\CGFccby

\usepackage[T1]{fontenc}
\usepackage{dfadobe}  

\usepackage{cite}  
\BibtexOrBiblatex

\electronicVersion
\PrintedOrElectronic
\ifpdf \usepackage[pdftex]{graphicx} \pdfcompresslevel=9
\else \usepackage[dvips]{graphicx} \fi

\usepackage{egweblnk} 
\usepackage[percent]{overpic}
\usepackage{caption}
\usepackage{graphicx}
\usepackage{amsfonts}
\usepackage{amsmath}
\usepackage{booktabs}
\usepackage{xcolor}
\usepackage{colortbl}
\usepackage{subcaption}
\usepackage{colortbl}

\definecolor{headerblue}{rgb}{0.8,0.87,0.94}
\definecolor{rowgray}{rgb}{0.95,0.95,0.95}

\usepackage{algorithm}
\usepackage{algorithmic}

\title[Swept Volume Computation]%
      {Swept Volume Computation with Enhanced Geometric Detail Preservation}

\author[Pengfei Wang \& Yuexin Yang \& Shuangmin Chen \& Shiqing Xin \& Changhe Tu \& Wenping Wang]
{\parbox{\textwidth}{\centering 
Pengfei Wang$^{*1}$\orcid{0000-0002-2079-275X},
Yuexin Yang$^{*1}$\orcid{0009-0009-4347-4517},
Shuangmin Chen$^{\dagger 2,3}$\orcid{0000-0002-0835-3316},
Shiqing Xin$^{1}$\orcid{0000-0001-8452-8723},
Changhe Tu$^{1}$\orcid{0000-0002-1231-3392},
Wenping Wang$^{4}$\orcid{0000-0002-2284-3952}
}
        \\
{\parbox{\textwidth}{\centering $^1$Shandong University~~~
         $^2$Qingdao University of Science and Technology~~~
         $^3$Shandong Key Laboratory of Deep Sea Equipment Intelligent Networking~~~
         $^4$Texas A\&M University\\
         $^*$Equal Contribution;~~~
         $^{\dagger}$ Corresponding Author.         
       }
}
}

\begin{document}

\definecolor{AAA}{rgb}{1.0, 0.13, 0.32}
\definecolor{BBB}{rgb}{0.2, 0.1, 1}
\definecolor{ccc}{rgb}{0, 0.5, 0.25}
\definecolor{DDD}{rgb}{0.0, 0.0, 0.0}
\captionsetup{labelfont=bf,textfont=it}
\newcommand{\PF}[1]{{\color{AAA}#1}}
\newcommand{\CG}[1]{{\color{DDD}#1}}
\newcommand{\SQ}[1]{{\color{AAA}SQ: #1}}
\newcommand{\YX}[1]{{\color{ccc}#1}}

\maketitle
\begin{abstract}
Swept volume computation—the determination of regions occupied by moving objects—is essential in graphics, robotics, and manufacturing. Existing approaches either explicitly track surfaces, suffering from robustness issues under complex interactions, or employ implicit representations that trade off geometric fidelity and face optimization difficulties. We propose a novel inversion of motion perspective: rather than tracking object motion, we fix the object and trace spatial points backward in time, reducing complex trajectories to efficiently linearizable point motions. Based on this, we introduce a multi-field tetrahedral framework that maintains multiple distance fileds per element, preserving fine geometric details at trajectory intersections where single-field methods fail. Our method robustly computes swept volumes for diverse motions, including translations and screw motions, and enables practical applications in path planning and collision detection.

\ccsdesc[500]{Computing methodologies~Shape modeling}

\printccsdesc   
\end{abstract}  
\section{INTRODUCTION}

\begin{figure}[htbp]
    \centering
     \begin{overpic}[width=0.7\columnwidth]{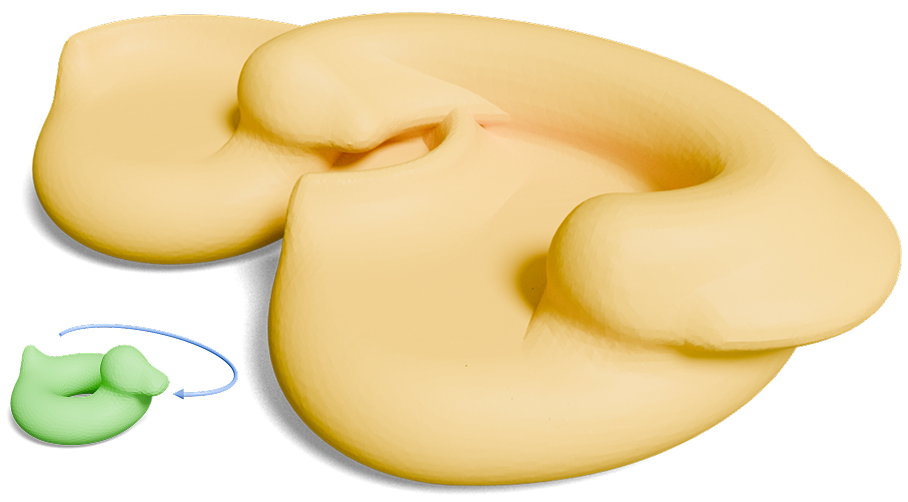}
    \end{overpic}
    \caption{Swept volume visualization. This geometric entity is formed by the union of all spatial positions occupied by an object as it moves along a defined trajectory, creating a continuous volumetric region. }
    \label{fig:sweepVolumeExample}
    \vspace{-2mm}
\end{figure}

\begin{figure}[htbp]
\centering
\begin{overpic}[width=1.0\columnwidth]{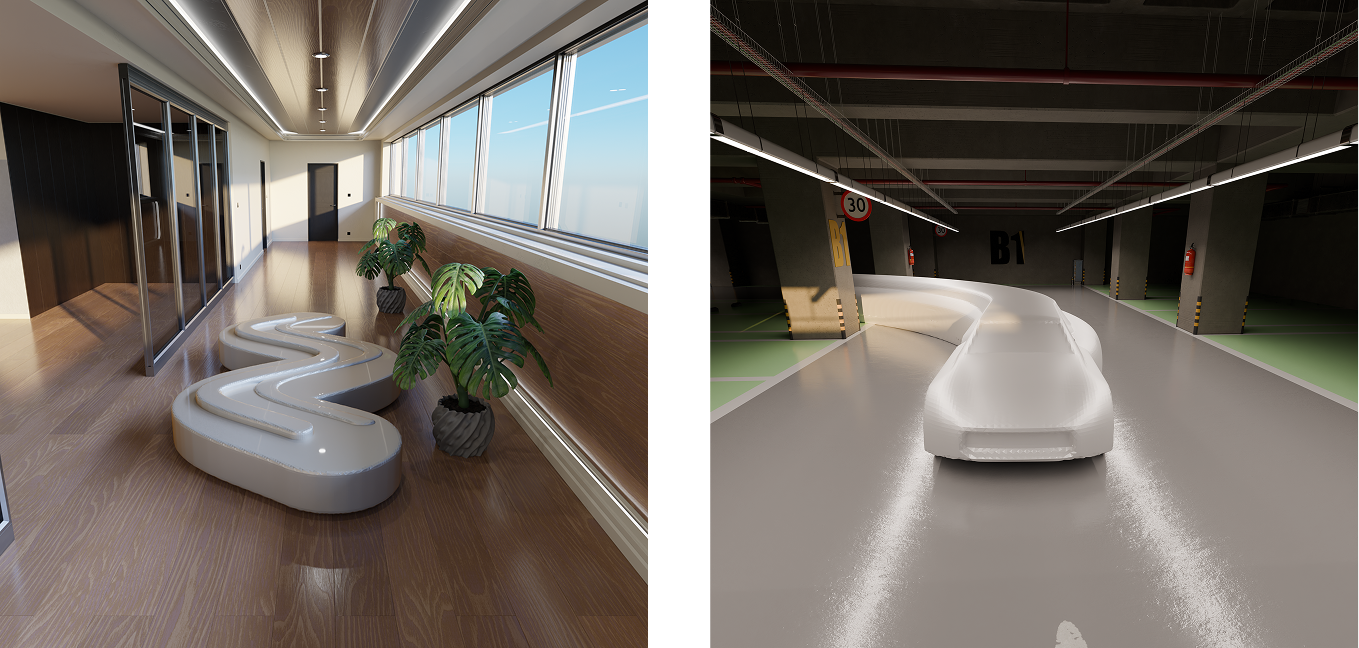}
\put(23,-4){(a)}
\put(74,-4){(b)}
\end{overpic}
\caption{Real-world examples: (a) robot vacuum navigation; (b) autonomous vehicle parking.}
\label{fig:car5}
\end{figure}

Swept volumes—the spatial regions traversed by moving objects—are fundamental geometric primitives in computer-aided design~\cite{SweptVolumes,DBLP:journals/corr/AdsulMS14,AbdelMalek2001OnSV,SHIH2005213}, robotics~\cite{abdel1997path,joho2024neuralimplicitsweptvolume,Murray2016RobotMP,Iwasaki2023OnlineMP}, and virtual reality~\cite{mcgraw2017interactive,Kim2007InteractiveCC}. Figure~\ref{fig:sweepVolumeExample} illustrates a swept volume generated by an object following a trajectory. 
\CG{And Figure~\ref{fig:car5} demonstrates two specific application scenarios: (a) robot vacuum path planning and collision avoidance, and (b) autonomous vehicle maneuvering in complex environments like underground parking facilities.}
Accurate swept volume computation significantly impacts collision detection~\cite{murray2016robot,Wang2024ImplicitSV,Xavier1997FastSD}, spatial reasoning~\cite{AbdelMalek2006SweptVF}, and geometric modeling~\cite{Laroche2020AnIR,Wang2000TheDE}. Despite their importance, efficiently generating high-quality swept volumes for complex shapes undergoing intricate motions remains challenging.

\CG{Traditional methods fall into two main categories: explicit geometry-based approaches that construct boundaries directly from meshes, and implicit field-based methods leveraging scalar fields.
Explicit geometry-based approaches directly operate on mesh representations to construct swept volume boundaries, ranging from analytical solutions for restricted motions (such as Minkowski sums for translations) to sophisticated polyhedral sweeping algorithms for general rigid motions~\cite{10.5555/194393,10.5555/6647,10.1177/027836499000900507}. These methods face fundamental challenges in robustly handling complex surface patch interactions during arbitrary motion.
Implicit field-based approaches represent swept volumes using scalar fields, typically signed distance functions (SDFs), to avoid explicit tracking of surface intersections~\cite{489382,346339,10.1145/3450626.3459780}.
Recent implicit approaches have made significant advances by attempting to obtain distance values from arbitrary spatial points to the resulting swept volumes through optimization. However, they are constrained by the complexity of input model SDFs and motion functions. 
The optimization process often converges slowly, becomes trapped in local minima, and lacks theoretical guarantees, especially for motions containing high-frequency details. Furthermore, traditional isosurface extraction strategies applied to the resulting SDFs struggle to preserve complex geometric details.
}

\CG{In this work, drawing inspiration from inverse analysis approaches in ray tracing methods~\cite{10.1145/15922.15918} and existing swept volume computation techniques~\cite{10.1145/3450626.3459780}, and addressing the limitations of single-field extraction methods, we propose a fundamentally different approach based on motion perspective inversion.} 
\CG{Our method takes mesh models as both input and output. }
Rather than tracking objects through space, we fix the object and consider spatial points traversing inverse trajectories through time.
This perspective shift transforms the challenging problem of analyzing complex object motion into the simpler
task of studying individual point trajectories, which can be linearly
approximated within sufficiently small temporal intervals.
Leveraging this insight, we introduce a multi-field isosurface extraction method maintaining multiple scalar fields within each tetrahedral element, capturing complex geometric features missed by single-field methods.

Our contributions include:
\begin{itemize}
\item \textbf{Motion perspective inversion}: Transforming complex motion tracking into tractable point-wise operations.
\item \textbf{Multiple distance field representation}: Preserving detailed geometric features via multiple scalar fields per tetrahedron.
\item \textbf{Four-dimensional incremental cutting}: Robustly extracting detailed swept volume surfaces.
\end{itemize}

These innovations enable accurate, efficient swept volume computations suitable for practical applications.

\section{RELATED WORK}

Swept volume computation methods have evolved primarily along two trajectories: explicit geometry-based approaches, which directly manipulate mesh representations, and implicit field-based approaches, which utilize scalar fields to represent swept regions. We review key developments in both categories, highlighting their fundamental principles, advantages, and limitations.

\subsection{Explicit Geometry-Based Methods}

Explicit methods operate directly on geometric representations, typically triangle meshes, to construct swept volume boundaries. Analytical solutions exist for restricted motion types, notably the Minkowski sum for pure translations, where Kaul et al.~\cite{10.5555/194393} demonstrated efficient computation of exact swept volumes for straight-line paths. However, this approach does not accommodate rotations or general spatial trajectories.

Early specialized approaches addressed specific motion types with varying degrees of success. Korein et al.~\cite{10.5555/6647} developed techniques for single-axis rotations by identifying extreme vertex trajectories and constructing piecewise-planar patches. Weld et al.~\cite{10.1177/027836499000900507} established the foundational theoretical insight that the union of all face-sweeps equals the complete swept volume. This principle underpins many subsequent approaches but leaves unresolved the challenge of efficiently determining which patches contribute to the outer boundary.

For general rigid motions, Abrams et al.~\cite{ComputingSweptVolumes} introduced a robust polyhedral sweeping algorithm that directly constructs boundary representations. Their method approximates edge sweeps as ruled surfaces by sampling the motion at discrete time steps and triangulating the resulting patches. These patches are combined with face instances at critical times, and a clipped arrangement is computed to extract the outer boundary. Intelligent heuristics reduce computational complexity by skipping interior edges. Although effective for moderate model complexities, this approach still requires careful handling of geometric intersections and Boolean operations.

Campen et al.~\cite{Campen2010PolygonalBE} advanced explicit methods through adaptive discretization, subdividing arbitrary motions into piecewise-linear segments and applying intelligent culling to eliminate redundant surface pieces. Their method handles complex motions more effectively than previous approaches, though combinatorial challenges remain significant in worst-case scenarios. More recently, Von et al.~\cite{6224921} employed compressed voxelizations and Delaunay refinement to enhance computational efficiency, particularly for industrial geometries.

Despite these advancements, explicit methods consistently face fundamental challenges: robustly managing complex surface interactions during arbitrary motion typically involves restrictive assumptions or computational approximations that can compromise accuracy. As motion complexity increases, the combinatorial explosion of potential feature interactions makes exact boundary calculation increasingly difficult.

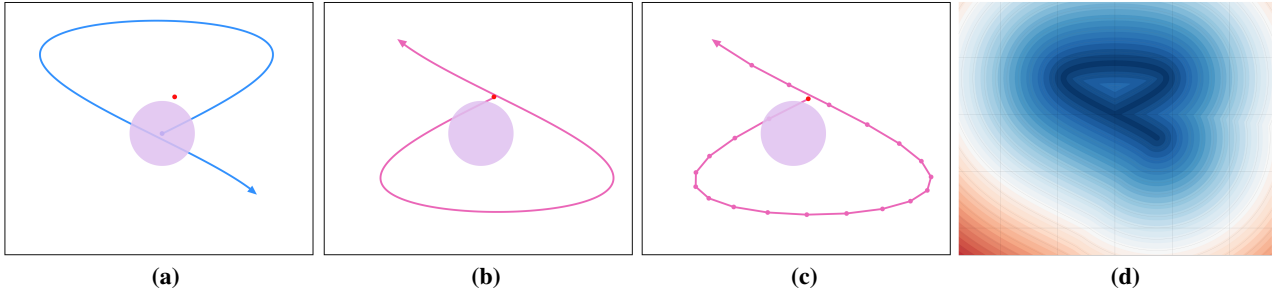
\begin{figure*}[htbp]
\begin{center}
    \begin{overpic}[width=0.95\linewidth]{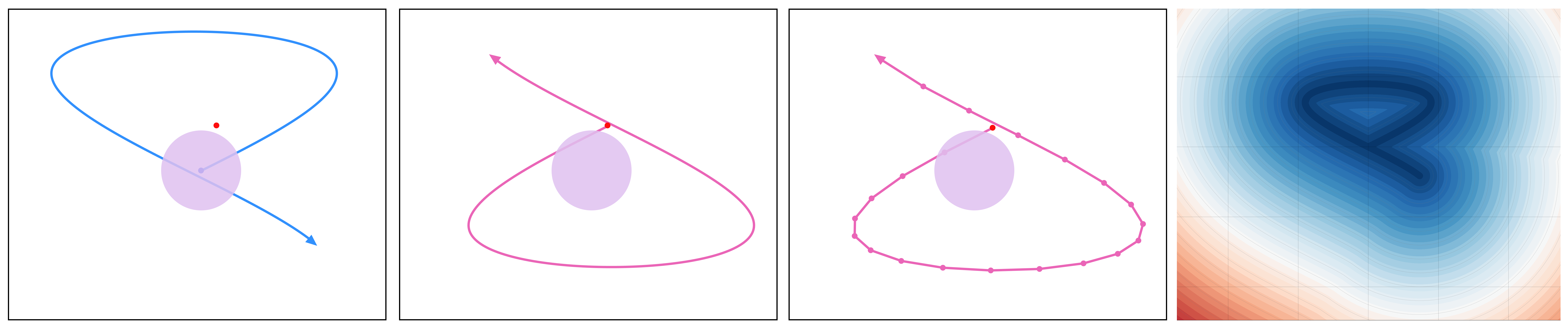}
        \put(12,-2){\textbf{(a)}}
        \put(37,-2){\textbf{(b)}}
        \put(62,-2){\textbf{(c)}}
        \put(87,-2){\textbf{(d)}}
    \end{overpic}
    \vspace{0.5em}  
     \caption{\CG{Distance Computation. (a) Given a model (purple circle as an example) and its sweeping trajectory, if we could compute the distance from any spatial point (exemplified by the red point) to the resulting swept volume, we would be able to construct the distance field. However, this is extremely challenging without the explicit swept volume result. 
     (b) Exploiting the relativity of motion, we invert the motion perspective: the model remains stationary while the query point traces an inverse trajectory. The red point lies inside the swept volume if and only if some point along its inverse trajectory lies inside the static model. The distance from the red point to the swept volume can be determined through inverse analysis of trajectory-to-model distances. (c) For complex motions, we partition the inverse trajectory temporally and assume linear motion within each small interval, enabling efficient computation through inverse analysis of segment-to-model distances as detailed in Section~\ref{sec:DistanceComputationviaInverseMotion} (d) Based on inverse motion analysis and linear approximation, we obtain the distance field for the swept volume. Note that this illustration shows a single distance field for demonstration; in practice, we employ a multi-field strategy where each distance field corresponds to a swept volume from a short time interval.}}
    \label{fig:overview_of_distance_computation}
\end{center}
\end{figure*}

\subsection{Implicit Field-Based Methods}

Implicit methods represent swept volumes using scalar fields, typically signed distance functions (SDFs), to avoid explicitly tracking surface intersections. In this paradigm, the swept volume comprises regions where the distance to the moving object becomes non-positive at any point during the motion.

Sourin et al.~\cite{489382} pioneered functional representations of swept volumes in 4D space-time. Their approach treats moving objects as time-dependent implicit functions and constructs swept volumes either by analytically uniting shape instances or by solving directly for the envelope condition. While elegantly handling arbitrary motions and even deforming objects, this formulation incurs significant computational costs.

A more practical implementation involves discretized sampling. Schroeder et al.~\cite{346339} converted moving objects into distance fields on 3D grids, progressively updating these fields as objects move to accumulate the swept volume. This ``voxel stamping'' method robustly manages self-intersections but suffers from resolution-dependent artifacts and high memory demands. Researchers have mitigated these limitations through adaptive sampling and hierarchical data structures, enhancing efficiency while retaining the robustness advantages of field-based approaches.

The most significant recent advancement in implicit methods comes from~\cite{10.1145/3450626.3459780}, who developed a numerical continuation approach in 4D space-time. Their method represents both the moving object and its swept volume as SDFs, tracing the swept surface as connected solution curves in spacetime. However, these 4D spacetime methods fundamentally require implicit function representations as input, significantly limiting their practical use in domains where explicit geometric representations dominate. Furthermore, computing distances to swept volumes of complex geometries undergoing intricate motions can encounter optimization difficulties, such as susceptibility to convergence to local minima, which undermines theoretical correctness guarantees.

\section{OVERVIEW}

Our method addresses two fundamental challenges in swept volume computation: (1) efficiently calculating the distance from a point to the swept volume, and (2) accurately preserving fine geometric details that emerge during the sweeping process.

To tackle the distance computation challenge, we invert the conventional perspective on motion analysis. Instead of tracking objects as they move through space, we keep the object fixed and analyze the inverse trajectories of spatial points. This transformation simplifies the complex problem of computing distances to swept volumes into the more manageable task of measuring distances between a static object and individual point trajectories, thereby providing a robust foundation for our extraction method.
\CG{Figure~\ref{fig:overview_of_distance_computation} illustrates this concept through a 2D example for intuitive understanding.}

To address the challenge of detail preservation, we employ tetrahedral discretization combined with a multi-field representation strategy. Unlike traditional methods that maintain a single scalar field within each element, our approach preserves multiple distance fields per tetrahedron—each corresponding to a swept volume segment generated during a specific time interval. To extract the final swept surface from these multiple fields, we introduce four-dimensional incremental cutting, processing the lower envelope of these fields in higher-dimensional space. This multi-field approach effectively captures intricate geometric features at trajectory intersections, overcoming the limitations inherent in single-field methods.

Section~\ref{sec:method} presents our two core algorithms: distance computation via inverse motion and isosurface extraction using four-dimensional incremental cutting. Section~\ref{sec:implementation} details the full implementation pipeline, demonstrating how these components effectively balance computational efficiency with geometric precision.

\begin{figure}[htbp]
\centering
\begin{overpic}[width=0.95\columnwidth]{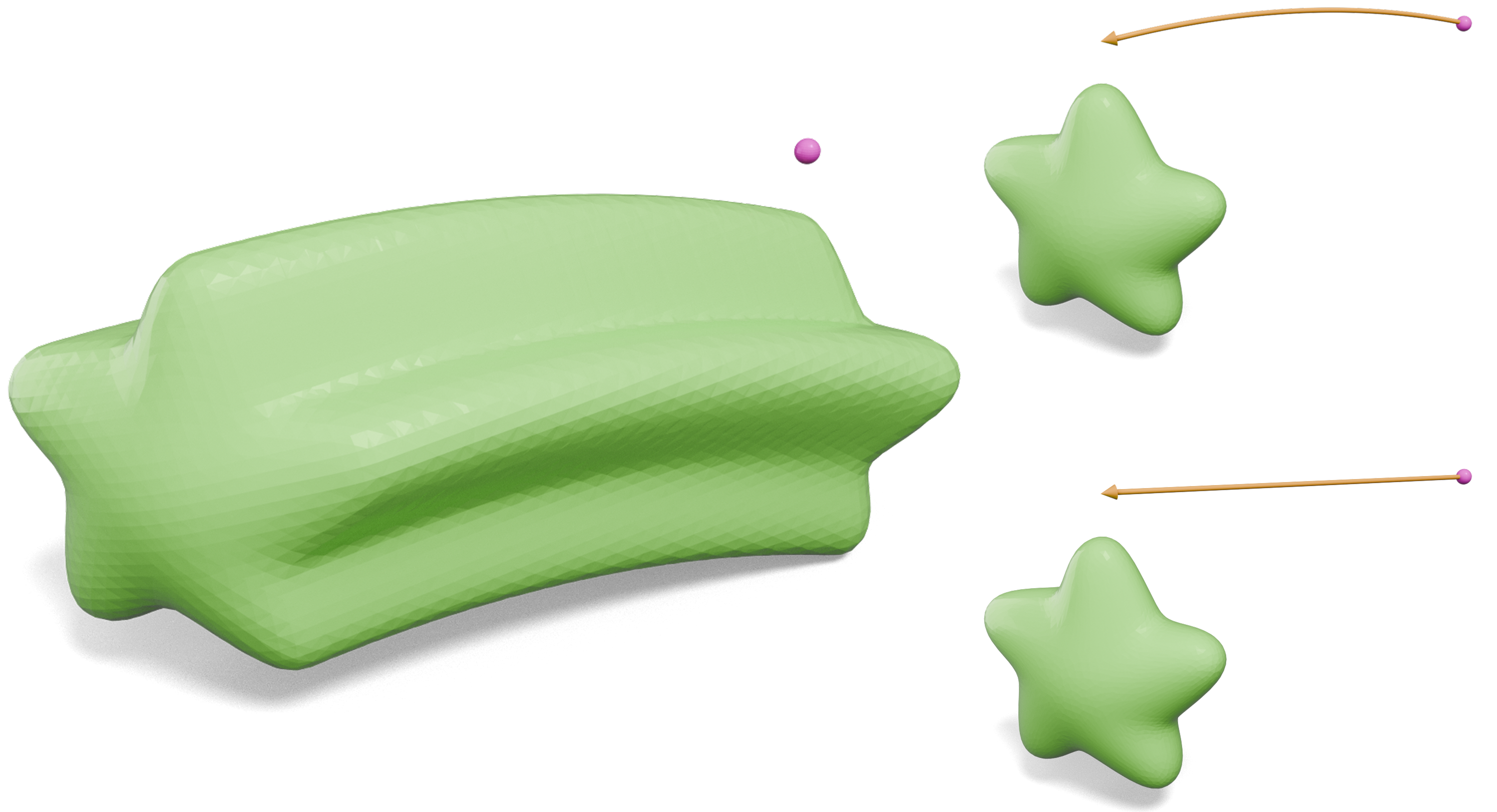}
\put(9,5){\textbf{(a)}}
\put(18.5,5){\textbf{$t_0$}}
\put(57,13.5){\textbf{$t_1$}}
\put(80,27){\textbf{(b)}}
\put(73,53.5){\textbf{$t_0$}}
\put(97,55){\textbf{$t_1$}}
\put(80,-3){\textbf{(c)}}
\put(73,23){\textbf{$t_0$}}
\put(97,24.5){\textbf{$t_1$}}
\end{overpic}

\caption{
\CG{Illustration of our relative motion analysis for short time intervals.} (a) Generate the swept volume by moving the object along the specified trajectory. (b) Keep the object stationary and move the observation point~$p$ along the reverse trajectory. (c) Approximate the short, curved trajectory with a straight-line segment. This reduces the determination of whether~$p$ lies within the swept volume to a simpler intersection test between the straight-line segment and the original stationary object.
\CG{Note that we illustrate the concept using simple motion within a short time interval here for demonstration purposes. For longer time intervals or complex motions, we subdivide the time period into finer segments and assume linear inverse trajectories within each segment, as shown in Figure~\ref{fig:timeDiscretization}.}}
\label{fig:segmentmove}
\end{figure}
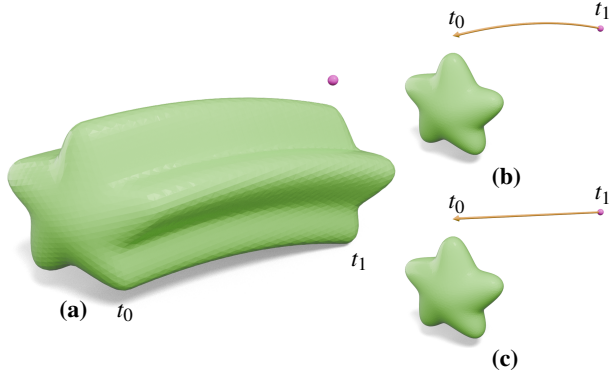

\section{METHOD}
\label{sec:method}

\subsection{Distance Computation via Inverse Motion}
\label{sec:DistanceComputationviaInverseMotion}
Computing the signed distance from an arbitrary point to the swept volume boundary constitutes a fundamental operation in swept volume extraction. 
However, given a short time interval $[t_0, t_1]$ and a query point $\mathbf{q}$ $\in$ $\mathbb{R}^3$, determining the distance from $\mathbf{q}$ to the swept volume of model $\mathcal{M}$ without explicitly constructing the swept volume presents significant computational challenges, as illustrated in Figure~\ref{fig:segmentmove}(a). 
\CG{In this subsection, we focus on scenarios with sufficiently short time intervals, while longer time intervals requiring piecewise linear approximation will be addressed in Section~\ref{sec:implementation}.}

\textbf{Motion Inversion.}
To determine the spatial relationship between a point and a swept volume
we invert the conventional perspective on motion to establish a more tractable computational framework.
Since motion is inherently relative, we can reformulate the problem by fixing the object's position and treating the static query point $\mathbf{q}$ as undergoing inverse motion, thus generating a trajectory through space (Figure~\ref{fig:segmentmove}(b)). Formally, if we define a continuous transformation function $\mathbf{f}(t): [t_0, t_1] \rightarrow SE(3)$ that maps time to a rigid transformation matrix, then the position of model $\mathcal{M}$ at time $t$ is given by $\mathcal{M}_t = \mathbf{f}(t)\mathcal{M}$.

Under our inverted perspective, the query point's trajectory can be expressed as:
\begin{equation}
\mathbf{q}(t) = \mathbf{f}^{-1}(t)\mathbf{q}, \quad t \in [t_0, t_1]
\end{equation}
A critical observation is that a point lies inside the swept volume if and only if its inverse trajectory intersects the static model—an exact correspondence that forms the theoretical foundation of our approach. 
\CG{Based on this principle, we can determine the distance from $\mathbf{q}$ to the swept volume of $\mathcal{M}$ during $[t_0, t_1]$ through distance analysis between the trajectory of $\mathbf{q}(t)$ and the static model $\mathcal{M}$.}

\textbf{Linear Trajectory Approximation.}
For sufficiently small time intervals $[t_0, t_1]$, the query point's motion can be approximated as linear, forming a line segment $L(t_0,t_1)$ connecting the positions at the interval endpoints:
\CG{
\begin{equation}
L(t_0,t_1) = \overline{\mathbf{q}(t_0)\mathbf{q}(t_1)} = \{\lambda\mathbf{q}(t_0) + (1-\lambda)\mathbf{q}(t_1) | \lambda \in [0, 1]\}
\end{equation}
}
As depicted in Figure~\ref{fig:segmentmove}(c), this simplification transforms our problem into computing the distance from line segment $L_{\mathbf{q}}$ to the static model $\mathcal{M}$:
\CG{
\begin{equation}
\label{equ:distanceComputation}
d(\mathbf{q}, \text{SweepVol}(\mathcal{M}, [t_0, t_1])) \approx \min_{p \in L(t_0,t_1)} d(p, \mathcal{M})
\end{equation}
}
\CG{where $d$ denotes the signed Euclidean distance.}

This transformation provides a computationally efficient approximation that preserves the essential geometric characteristics of the swept volume with high fidelity. The trajectory-based formulation significantly simplifies the computation process while maintaining the robustness necessary for accurate swept volume reconstruction.

\CG{
Note that we assume $[t_0, t_1]$ represents a very short time interval here, allowing the inverse trajectory to be approximated with a single line segment. In practice, the complete temporal motion is divided into multiple short intervals, with each inverse trajectory segment assumed to be linear, as detailed in Section~\ref{sec:SpatiotemporalDiscretization}.
}

\begin{figure}[htbp]
    \centering 
    \begin{overpic}[width=0.95\columnwidth]{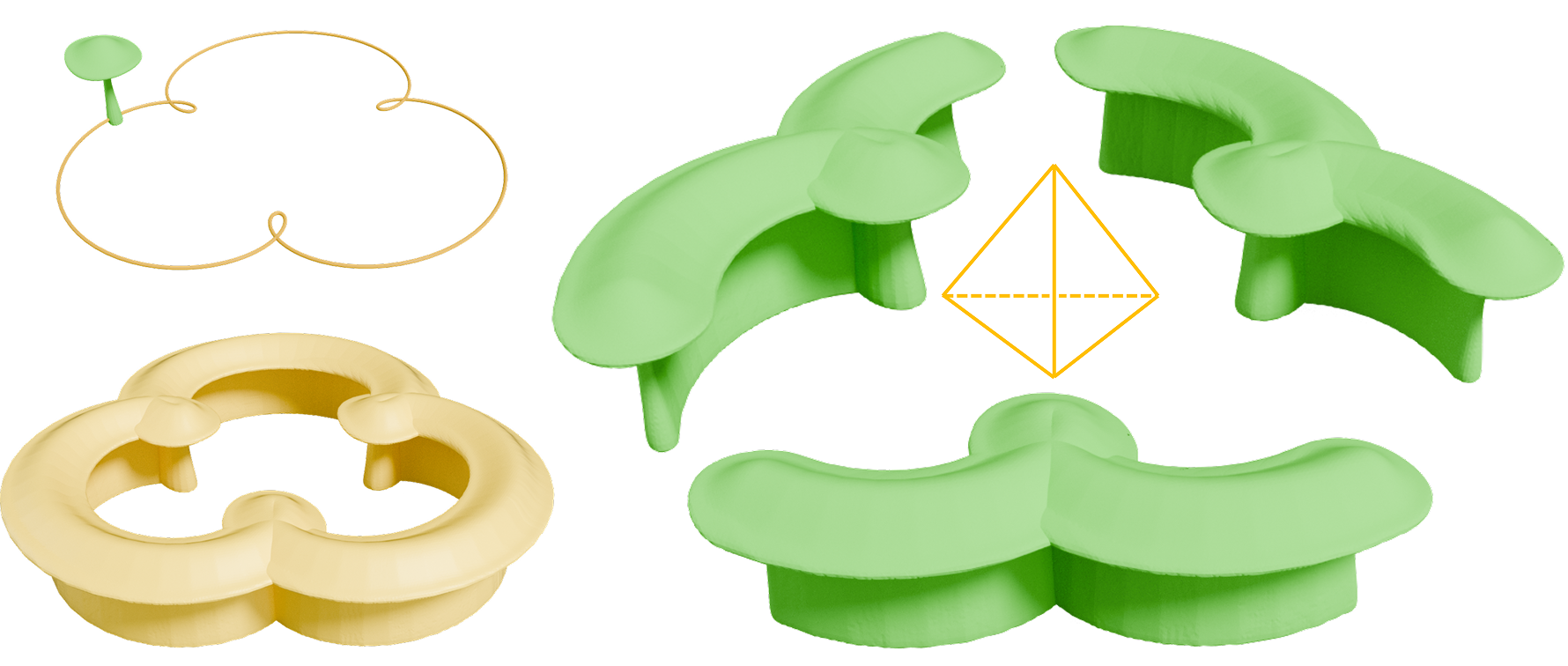}
    \put(-1,40){\textbf{(a)}}
    \put(-1,20){\textbf{(b)}}
    \put(40,40){\textbf{(c)}}
    \end{overpic}
    \caption{Multiple distance field representation in our approach. (a) Input model and motion trajectory. (b) Complete swept volume. 
    \CG{(c) The entire motion is temporally partitioned into multiple segments, with each segment generating a partial swept volume. Each partial swept volume contributes a distance field, illustrated here with three segments providing three distinct distance fields to the tetrahedron in the discretization. In practice, finer temporal discretization is employed.}
    }
    \label{fig:timespa}
\end{figure}

\begin{figure}[htbp]
\begin{center}
    \begin{overpic}[width=1.0\linewidth]{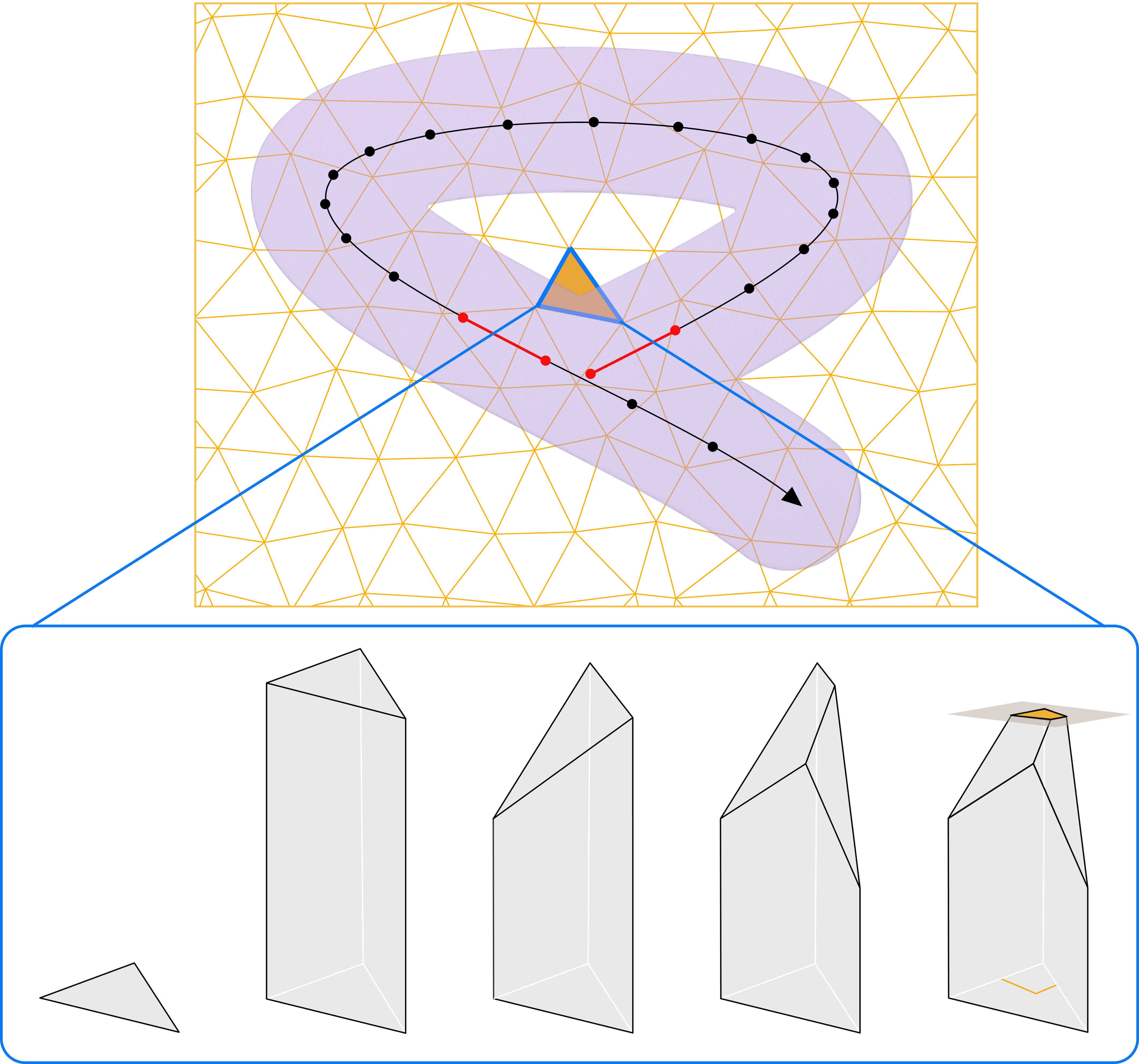}
        \put(8,1){\textbf{(a)}}
        \put(26,1){\textbf{(b)}}
        \put(48,1){\textbf{(c)}}
        \put(67,1){\textbf{(d)}}
        \put(87,1){\textbf{(e)}}
        \put(89,33){\textbf{iso=0}}
    \end{overpic}
     \caption{\CG{Feature preservation through multiple distance fields and incremental cutting in 2D swept volume extraction. Taking the highlighted triangle as an example, it stores two distance fields from swept volumes during different time segments, with trajectory segments from different time periods illustrated and contributing segments marked in red. (a) Original triangle with multiple distance fields defined within. (b) Unbounded triangular prism formed by extending the triangle infinitely along the $z$-axis. (c,d) Progressive incremental cutting operations using hyperplanes representing individual distance fields, where the height corresponds to distance field values, resulting in a lower envelope. (e) Intersecting the resulting prism with the $z=0$ plane and projecting the intersection curve back to 2D space to obtain the zero-contour within the triangle.}}
    \label{fig:example}
\end{center}
\end{figure}
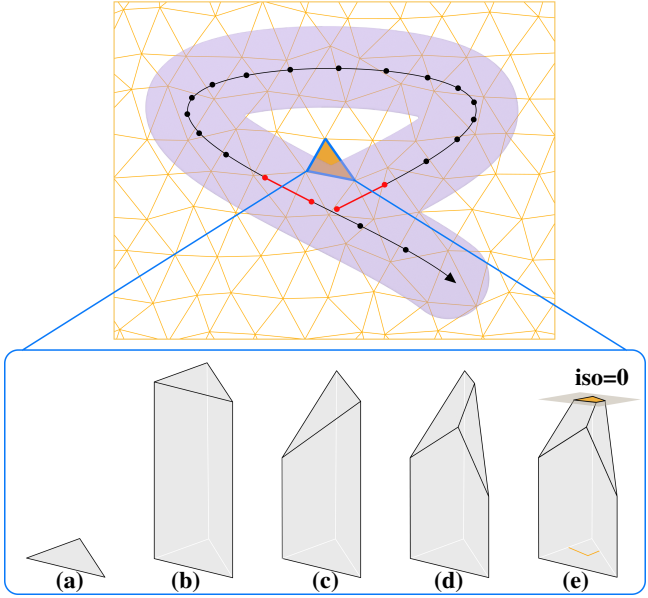

\subsection{4D Incremental Cutting for Isosurface Extraction}
\label{sec:4DIncrementalCuttingforIsosurfaceExtraction}
\CG{
Isosurface extraction from implicit fields is a fundamental task in computational geometry~\cite{Wyvill,10.1145/2732197}. 
}
Traditional isosurface extraction methods such as Marching Cubes~\cite{10.1145/37401.37422} and Dual Contouring~\cite{10.1145/566654.566586} operate on scalar fields where a single distance value is stored at each vertex of a spatial tessellation. These approaches process each volumetric element where the zero-isosurface passes through, using the distance values at element vertices (and potentially gradient information) to reconstruct surface fragments. However, a significant limitation of these conventional methods is their fundamental assumption of surface simplicity within each element, which renders them inadequate for capturing complex geometric features and intricate surface structures that emerge in swept volumes, particularly at trajectory intersections.

To overcome this limitation, our method employs tetrahedral tessellation of the computational domain and, critically, maintains multiple distance fields within each tetrahedron.
Each of these fields corresponds to the signed distance field of a swept volume generated by model $\mathcal{M}$ during a different time interval of its motion.
As shown in Figure~\ref{fig:timespa}, the complete trajectory is divided into three segments, 
with the swept volume formed by each trajectory segment providing a separate distance field for the tetrahedron.

Let us denote the swept volume of model $\mathcal{M}$ during time interval $[t^s_i, t^e_i]$ as $\text{SV}_i(\mathcal{M})$.
For a tetrahedron $t$, we denote the set of distance fields as $\{\pi_i\}_{i=1}^m$, where each $\pi_i$ corresponds to the signed distance field to $\text{SV}_i(\mathcal{M})$.

Following the approach introduced in~\cite{10845125,10.1145/3550454.3555453}, we assume linear variation of each distance field within the tetrahedron $t$.
Specifically, each distance field $\pi_i$ is encoded by four values $\pi_i = \{d_{1,i}, d_{2,i}, d_{3,i}, d_{4,i}\}$, where $d_{k,i}$ represents the signed distance from the $k$-th vertex of the tetrahedron to $\text{SV}_i(\mathcal{M})$. For any point $x$ within $t$ with barycentric coordinates $(v_1, v_2, v_3, v_4)$, the signed distance from $x$ to $\text{SV}_i(\mathcal{M})$ can be approximated through linear interpolation:
\begin{equation}
\pi_i(x) = v_1 d_{1,i} + v_2 d_{2,i} + v_3 d_{3,i} + v_4 d_{4,i}
\end{equation}

The actual distance field within $t$, accounting for all component fields, is determined by taking the minimum value across all distance fields:
\begin{equation}
d(x) = \min_{i \in \{1, 2, ..., m\}} \pi_i(x)
\end{equation}

Our goal is to compute the zero-isosurface of this combined distance field.
\CG{For easier understanding, we consider a simplified 2D case. Figure~\ref{fig:example} illustrates this concept through a 2D example for intuitive understanding, where the motion time is divided into multiple segments, with each segment's swept volume generating a distance field computed using the approach shown in Figure~\ref{fig:segmentmove}. In this scenario, multiple distance fields exist within the highlighted triangle, with each field visualized as a plane in 3D space where the $z$-axis represents the interpolated distance value at each point. We perform incremental linear cutting operations on an unbounded triangular prism (whose base is the original triangle, extending infinitely upward and downward) using these planes, continuously updating the lower envelope. The zero-contour is then obtained by intersecting this lower envelope with the $z=0$ plane and projecting the resulting intersection curve back to the 2D domain.}

Extending this approach to 3D, we implement incremental cutting in 4D space to extract the zero-isosurface of the distance field, following the approach described in~\cite{10845125}. Each tetrahedron, along with its associated distance fields, defines a set of hyperplanes in 4D space. By computing the lower envelope of these hyperplanes and intersecting it with the zero-hyperplane (where the distance equals zero), then projecting the result back to 3D space, we obtain the zero-isosurface structure within the tetrahedron. This 4D incremental cutting approach faithfully captures complex geometric features generated during the object's motion, including sharp features and intricate surface structures that would otherwise be lost with traditional isosurface extraction methods.
\CG{Algorithm~\ref{alg:isosurface-extraction} presents the pseudocode.}

\begin{algorithm}[ht]
\CG{
\caption{Isosurface Extraction via 4D Incremental Cutting}
\label{alg:isosurface-extraction}
\begin{algorithmic}[1]
\REQUIRE Tetrahedron $\tau$, Distance fields $\{\pi_1, \pi_2, \ldots, \pi_m\}$ stored in $\tau$
\ENSURE Swept volume surface within tetrahedron $\tau$

\STATE Initialize 4D prismatic structure $P$ with base $\tau$ and infinite height range $(-\infty, +\infty)$ in $w$-dimension
\FOR{each distance field $\pi_i$ in $\{\pi_1, \pi_2, \ldots, \pi_m\}$}
    \STATE Construct 4D hyperplane $H_i$ from distance values of $\pi_i$ at tetrahedron vertices
    \STATE Cut $P$ by hyperplane $H_i$ and retain the portion below $H_i$ (lower envelope)
\ENDFOR
\STATE Intersect the final lower envelope structure $P$ with hyperplane $w = 0$
\STATE Extract intersection surface $S$ between $w = 0$ hyperplane and the lower envelope
\STATE Project surface $S$ back to 3D space to obtain swept volume boundary within $\tau$
\RETURN Swept volume surface mesh in $\tau$
\end{algorithmic}
}
\end{algorithm}

\section{IMPLEMENTATION}
\label{sec:implementation}
Our implementation consists of three phases:
\begin{itemize}
    \item \textbf{Spatiotemporal Discretization}: We partition the spatial domain into tetrahedral elements and divide the motion into small time intervals that support our linear trajectory approximation.
    \item \textbf{Distance Field Propagation}: For each time interval, we propagate distance fields from seed tetrahedra throughout the tetrahedral discretization  using a competition mechanism, ensuring each tetrahedron receives the necessary fields for isosurface extraction.
    \item \textbf{Swept Volume Extraction}: We apply four-dimensional incremental cutting to each tetrahedron independently, converting multiple distance fields into a coherent swept volume boundary.
\end{itemize}
We now describe each component of this pipeline in detail.

\begin{figure}[htbp]
    \centering
     \begin{overpic}[width=0.95\columnwidth]{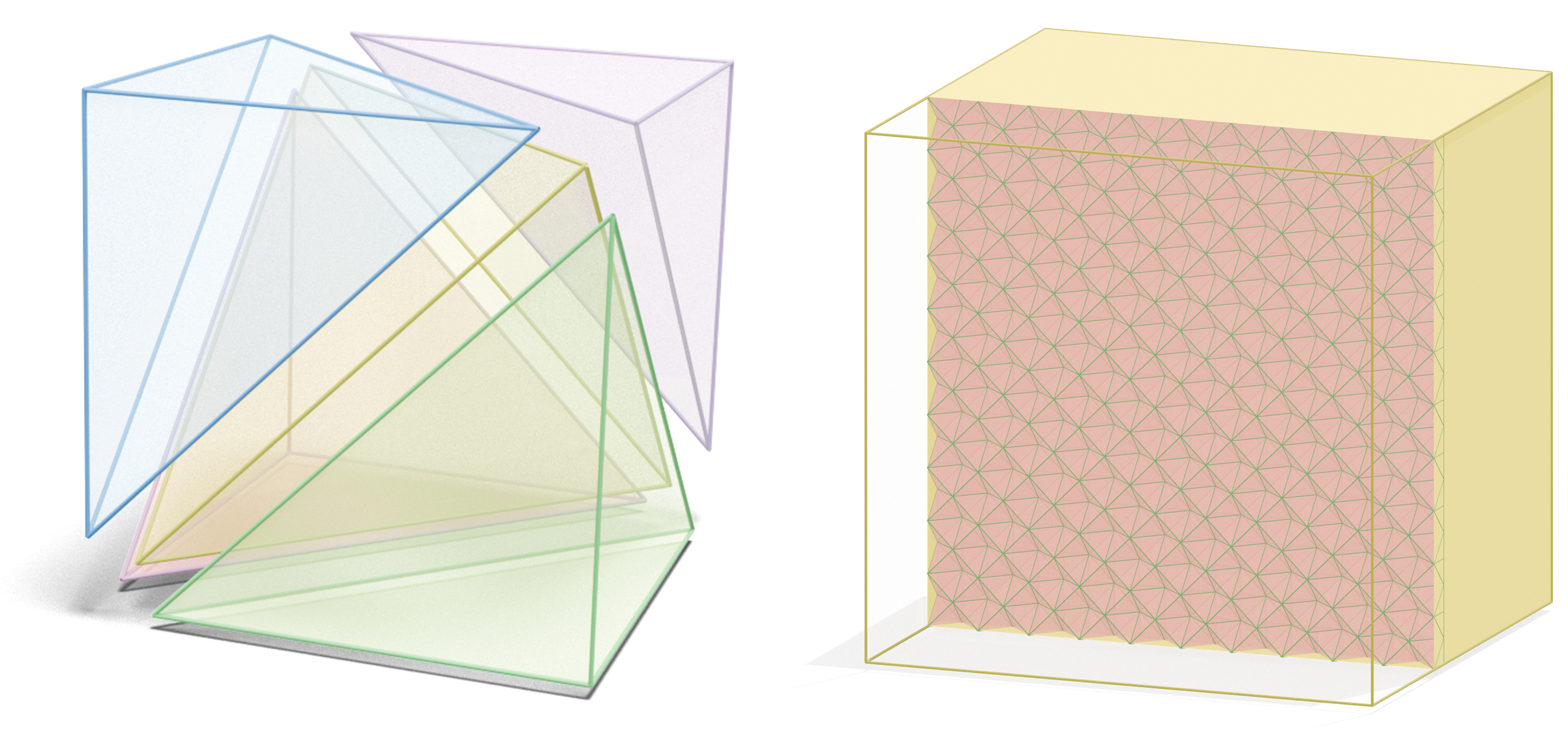}
      \put(20,-3){\textbf{(a)}}
      \put(70,-3){\textbf{(b)}}
    \end{overpic}
    \caption{Spatial discretization. (a) A cube subdivided can be into five tetrahedra. (b) Tetrahedral tessellation of the computational domain where we first partition space into uniform cubes, then subdivide each cube as shown in (a).}
    \label{fig:5tets}
\end{figure}

\subsection{Spatiotemporal Discretization}
\label{sec:SpatiotemporalDiscretization}
Our method employs discretization in both spatial and temporal dimensions to facilitate subsequent computational operations. This discretization strategy establishes the foundation for our swept volume computation framework.

\textbf{Spatial Discretization.}
We utilize tetrahedral tessellation to discretize the computational domain. Specifically, we first partition the solution space using uniform cubic elements, then subdivide each cube into five tetrahedra as illustrated in Figure~\ref{fig:5tets}(a). Through careful arrangement, we maintain spatial consistency across the tetrahedral mesh. Figure~\ref{fig:5tets}(b) presents an example of our tetrahedral tessellation applied to a computational domain.

\begin{figure}[htbp]
\begin{center}
    \begin{overpic}
    [width=0.6\columnwidth]{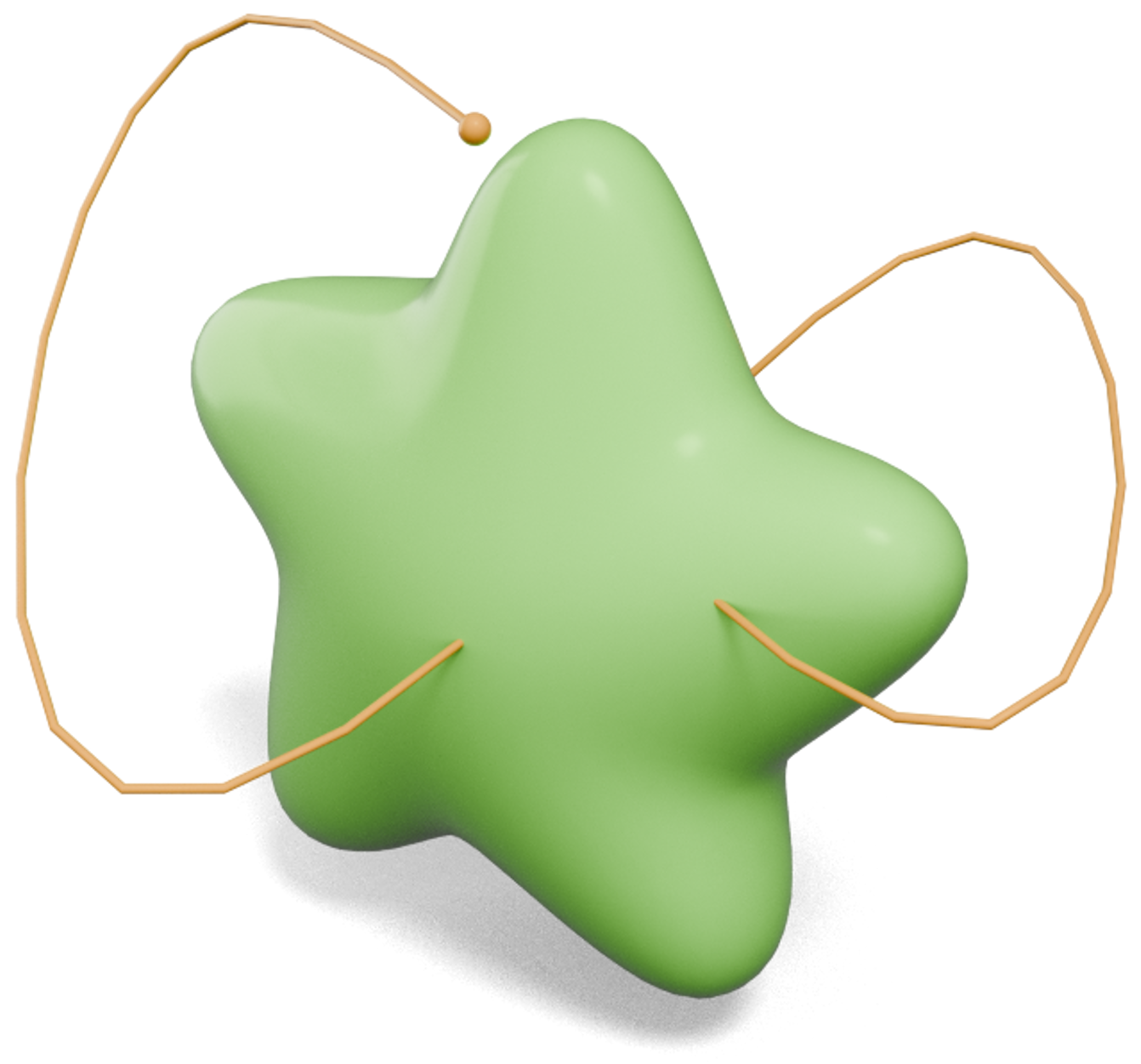}

    \end{overpic}
     \caption{Temporal discretization. The green model remains stationary while orange curves show inverse trajectories of spatial points. Within each discrete time interval, these trajectories are approximated as linear segments.}
    \label{fig:timeDiscretization}
\end{center}
\end{figure}

\textbf{Temporal Discretization.}
In the temporal dimension, we divide the entire motion time interval into $N$ segments. Within each temporal segment, we assume the inverse trajectories of spatial points can be well-approximated as linear segments, as shown in Figure~\ref{fig:timeDiscretization}. This piecewise linear approximation is consistent with the motion perspective inversion described in Section \ref{sec:DistanceComputationviaInverseMotion} and allows us to handle complex motion paths while maintaining computational efficiency. The resolution of this temporal discretization can be adjusted based on the complexity of the original motion and the desired accuracy of the final swept volume representation.

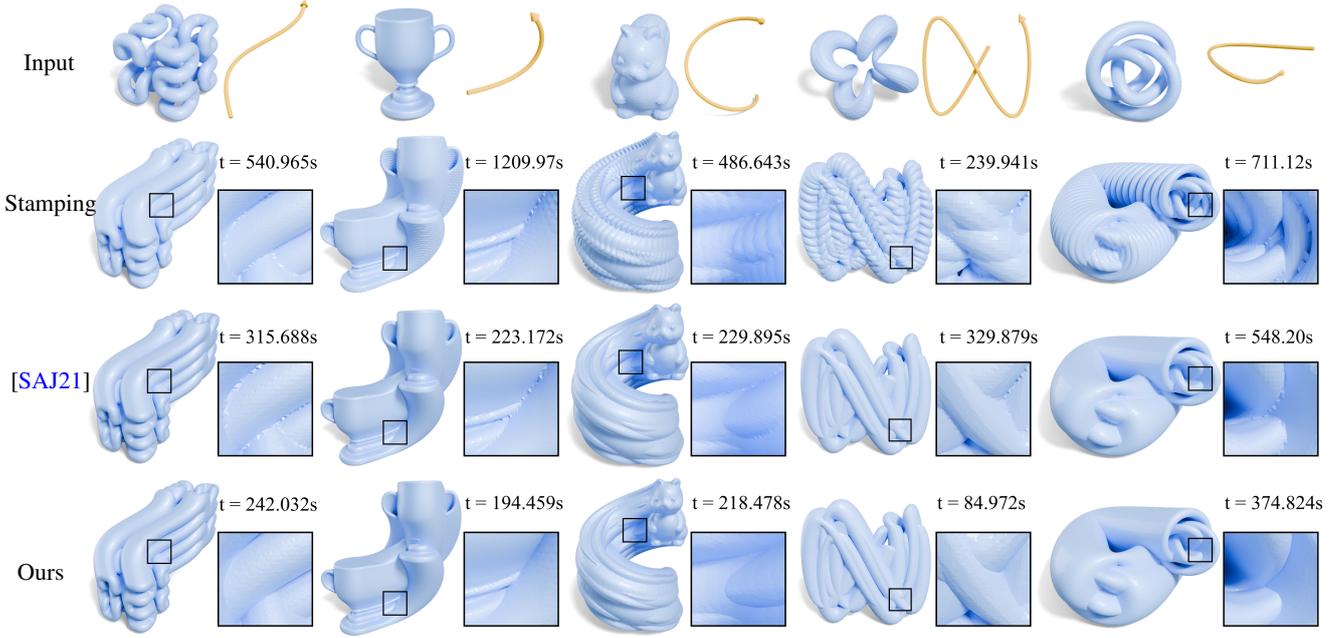
\begin{figure*}[htbp]
    \centering 
    \begin{overpic}[width=2.0\columnwidth]{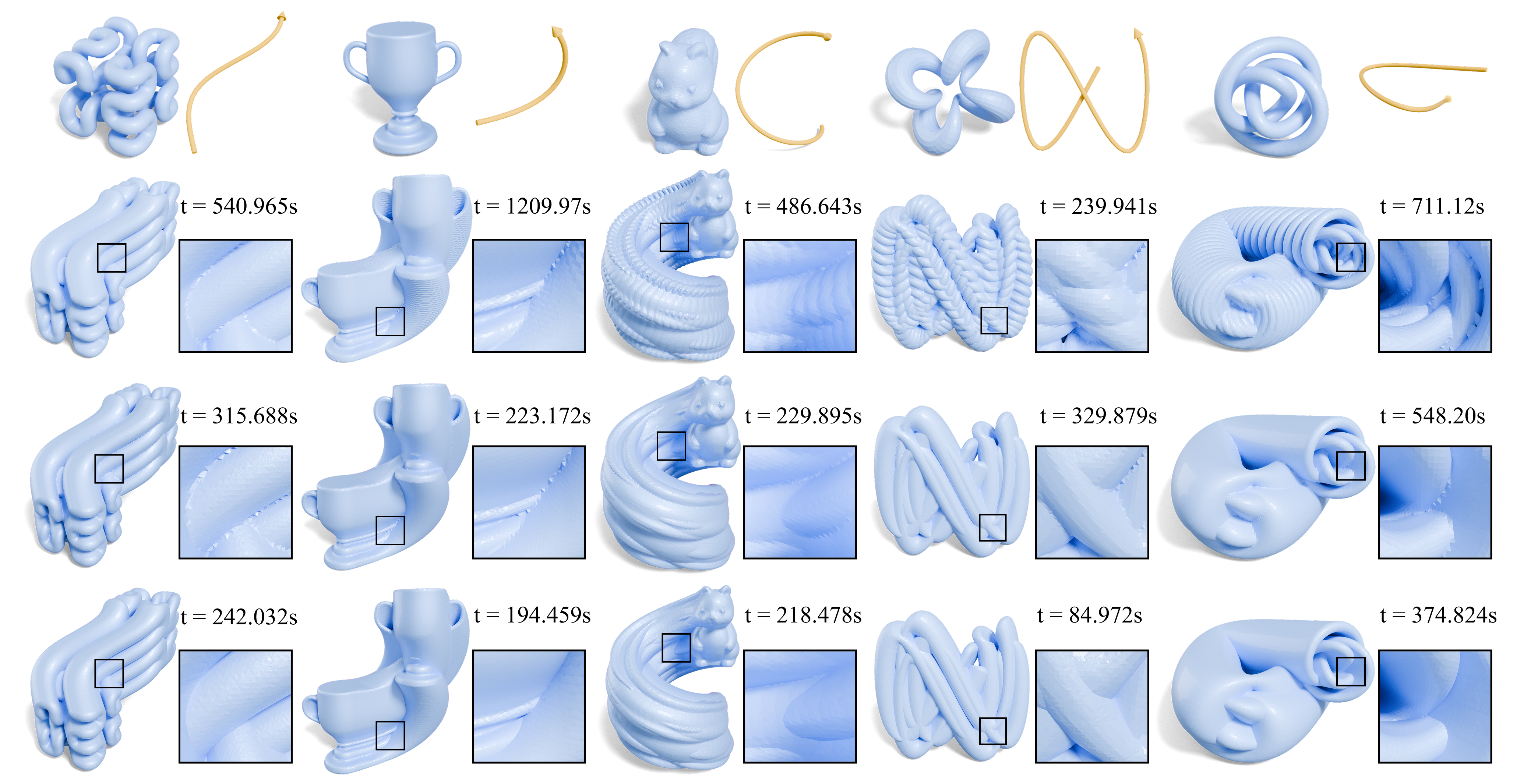}
    \put(-3.5,45){Input}
    \put(-5,34){Stamping}
    \put(-5,20){\cite{10.1145/3450626.3459780}}
    \put(-4,5){Ours}
    \end{overpic}
    \caption{Visual comparison of swept volumes generated by our method against Stamping and~\cite{10.1145/3450626.3459780} across different models and motion types. Our method preserves sharp features while avoiding the zigzag artifacts present in the Stamping approach.}
    \label{fig:compare_expirement}
\end{figure*}

\subsection{Spatial Propagation of Piecewise Linear Motion Distance Fields}

We consider a normalized motion time interval $[0, 1]$ discretized into $N$ segments, with each segment spanning $[t_i, t_{i+1}]$. For model $\mathcal{M}$, the swept volume formed during time interval $[t_i, t_{i+1}]$ is denoted as $\text{SV}(\mathcal{M}, t_i, t_{i+1})$. Each segment-specific swept volume generates a distance field throughout the computational domain, with the complete swept volume defined by taking the minimum value among all these fields at each point.

\begin{algorithm}[ht]
\CG{
\caption{Line Segment to Model Distance Query}
\label{alg:segment-distance}
\begin{algorithmic}[1]
\REQUIRE Line segment $L = \{p_0, p_1\}$, Model $\mathcal{M}$
\ENSURE Signed distance $d$

\IF{line segment $L$ lies completely outside model $\mathcal{M}$}
    \STATE $d \leftarrow$ distance from $L$ to $\mathcal{M}$ using FCPW library
\ELSE
    \STATE Extract the portion of $L$ that lies inside $\mathcal{M}$ as $L_{\text{interior}}$
    \STATE Sample 10 evenly distributed points along $L_{\text{interior}}$
    \STATE $d \leftarrow +\infty$
    \FOR{each sampled point $p$}
        \STATE Compute signed distance $d_p$ from $p$ to $\mathcal{M}$
        \STATE $d \leftarrow \min(d, d_p)$
    \ENDFOR
\ENDIF
\RETURN $d$
\end{algorithmic}
}
\end{algorithm}

\begin{algorithm}[ht]
\CG{
\caption{Distance Field Propagation for Swept Volume Computation}
\label{alg:distance-propagation}
\begin{algorithmic}[1]
\REQUIRE Tetrahedral mesh $\mathcal{T}$, Model $\mathcal{M}$, Global motion function $f(t)$, $N$ time intervals $[t_i, t_{i+1}]$
\ENSURE Distance fields stored in each tetrahedron

\STATE Initialize empty priority queue $Q$ for distance propagation
\FOR{each time interval $[t_i, t_{i+1}]$}
    \STATE Randomly select fixed number of seed tetrahedra inside $\mathcal{M}(t_i)$
    \FOR{each seed tetrahedron $\tau$}
        \STATE Add propagation event $(\tau, [t_i, t_{i+1}])$ to priority queue $Q$
    \ENDFOR
\ENDFOR

\WHILE{$Q$ is not empty}
    \STATE Extract top-priority event $(\tau, [t_i, t_{i+1}])$ from $Q$
    \STATE Compute distance field $\pi$ within $\tau$ for swept volume during time interval $[t_i, t_{i+1}]$ using motion inversion
    \IF{$\pi$ is not defeated by existing fields in $\tau$ and not all distance values of $\pi$ at vertices are positive}
        \STATE Store $\pi$ in the distance field list of $\tau$
        \FOR{each neighboring tetrahedron $\tau'$ of $\tau$}
            \STATE Add propagation event $(\tau', [t_i, t_{i+1}])$ to $Q$
        \ENDFOR
    \ENDIF
\ENDWHILE
\end{algorithmic}
}
\end{algorithm}
\CG{To compute distances from points to these swept volumes, we employ the trajectory-based approach described in Section~\ref{sec:DistanceComputationviaInverseMotion}, which reduces to analyzing the relationship between line segments and the static model.}
\CG{According to Equation~\ref{equ:distanceComputation}, when the line segment lies completely outside the model, the minimum distance from all points on the segment to the model equals the segment-to-model distance, which we compute using the FCPW library~\cite{FCPW}. However, when portions of the segment lie inside the model, we discretize the interior portion along the segment into sample points, compute the signed distance from each point to the model, and take the minimum signed distance. This discretization approach is necessary because we require the minimum signed distance rather than the minimum absolute distance, and existing libraries cannot provide the required computation when segments lie partially inside the model. Algorithm~\ref{alg:segment-distance} presents the pseudocode for line segment-to-model distance calculation.}
While this introduces some approximation in the absolute distance values, the sign of the distance remains correct, which is crucial for accurate inside/outside determination. 
In our experiments, we used 10 sample points for this discretization.

As described in Section~\ref{sec:4DIncrementalCuttingforIsosurfaceExtraction}, within each tetrahedron, we represent each distance field $\pi_i$ by four values corresponding to the signed distances at its vertices: $\pi_i = {d_{1,i}, d_{2,i}, d_{3,i}, d_{4,i}}$. To determine which distance fields contribute to each tetrahedron, we adopt the competition mechanism from prior work~\cite{10.1145/3550454.3555453}. For two distance fields $\pi_i$ and $\pi_j$ within tetrahedron $t$, we consider $\pi_j$ to be defeated by $\pi_i$ if: 
\begin{equation}
d_{1,i} < d_{1,j} \text{ and } d_{2,i} < d_{2,j} \text{ and } d_{3,i} < d_{3,j} \text{ and } d_{4,i} < d_{4,j}.
\end{equation}
Our goal is to identify all non-defeated distance fields for each \CG{Useful} tetrahedron, which collectively define the local swept volume boundary.

For each time interval $[t_i, t_{i+1}]$, we select a fixed number of seed tetrahedra that lie inside $\mathcal{M}(t)$ for some time $t$ within this interval and initialize them with the corresponding segment's distance field. 
\CG{These distance fields propagate to neighboring tetrahedra only if they remain undefeated in current tetrahedron and not all four distance values at the tetrahedron vertices are positive (as fields with all positive values contribute no meaningful information to swept volume surface extraction within that tetrahedron).}
Each tetrahedron stores all undefeated distance fields it receives during this process, which collectively define its contribution to the swept volume boundary. This propagation process is efficiently managed using a priority queue, and the algorithm terminates when no distance fields remain to be propagated. This approach ensures that all tetrahedra containing relevant portions of the swept volume receive their appropriate distance fields. The detailed implementation of this propagation strategy follows the approach described in~\cite{10.1145/3550454.3555453}.
\CG{
For better understanding, we provide the pseudocode for distance field propagation in Algorithm~\ref{alg:distance-propagation}.
}

\subsection{Isosurface Extraction}
After determining all contributing distance fields for each tetrahedron, we extract the swept volume boundary using the four-dimensional incremental cutting approach described in Section~\ref{sec:4DIncrementalCuttingforIsosurfaceExtraction}. This process converts the multiple distance field representations within each tetrahedron into a coherent surface representation of the swept volume boundary.

Our implementation of the 4D incremental cutting procedure follows the approach presented in~\cite{10845125}, adapted to the specific context of swept volume computation. A key advantage of our method is that isosurface extraction can be performed independently for each tetrahedron, without inter-element dependencies. This property makes the algorithm well-suited for parallel acceleration, enabling efficient execution on multi-core CPUs. In our implementation, we exploit this parallelism to substantially reduce computation time, particularly for complex models and motion paths where many tetrahedra contain non-trivial swept volume contributions.

\section{EVALUATION}

\subsection{Experimental Setting}

\textbf{Hardware Environment.}
Our algorithm was implemented in C++ on a computing platform equipped with a 3.4 GHz Intel Core i7-14700K 20-Core CPU, 64 GB of memory, and Windows 11 operating system.
\CG{We employed 28 threads for parallel acceleration during the isosurface extraction phase using the OpenMP framework.}
\CG{No post-processing smoothing was applied in our experiments.}

\noindent
\textbf{Baseline Methods.}
We compared our approach against two representative methods:
\begin{itemize}
\item \textbf{Stamping}: This method, commonly employed in applications such as Adobe Medium, involves sampling the moving object at discrete intervals. At each sample, a signed distance field (SDF) representation is generated. These SDFs are combined by taking the minimum value at each spatial location, followed by extracting the zero-isosurface to form the swept volume boundary.
\item \textbf{\cite{10.1145/3450626.3459780}}: This optimization-based approach calculates the minimal distance values at each spatial grid vertex throughout the sweeping process, utilizing dual contouring to extract the final swept volume surface.
\end{itemize}

Consistent discretization parameters were applied across all methods. The Spacetime Numerical Continuation method requires intermediate states for trajectory interpolation, while our method and the Stamping approach need temporal discretization. We uniformly adopted a sampling rate of $50$ time steps for all experiments. Surface extractions were performed at a spatial resolution of $256^3$.

Additionally, we initialized distance field computations from $100$ randomly selected seed tetrahedra within each relevant time interval, subsequently propagating fields throughout the spatial discretization.

\begin{figure*}[htbp]
    \centering 
    \begin{overpic}[width=2.1\columnwidth]{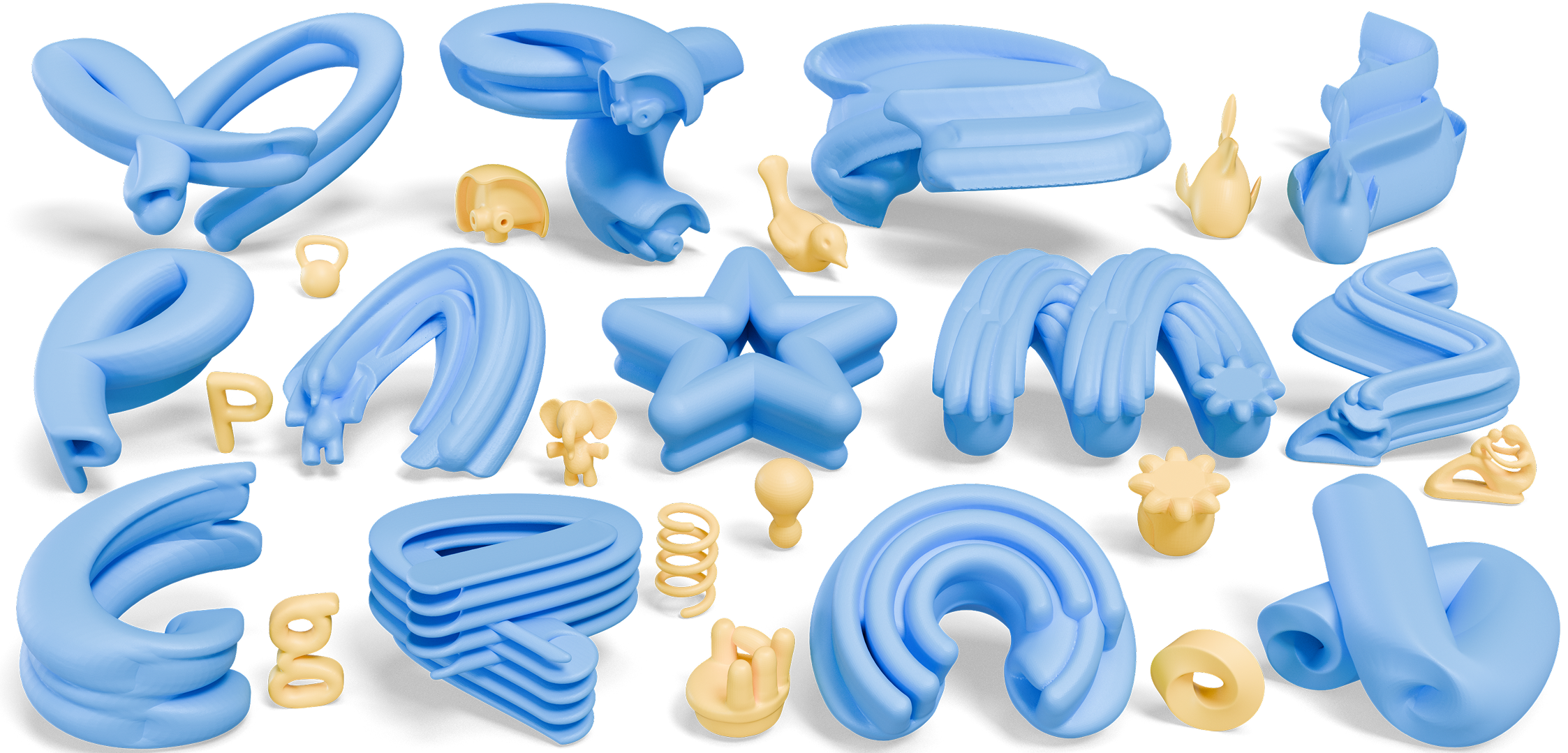}
    
    \end{overpic}
\caption{Diverse swept volume results generated by our algorithm for various geometric models undergoing different motion trajectories.}
    \label{fig:gallery}
\end{figure*}
\begin{figure}[htbp]
    \centering 
    \begin{overpic}[width=0.65\columnwidth]{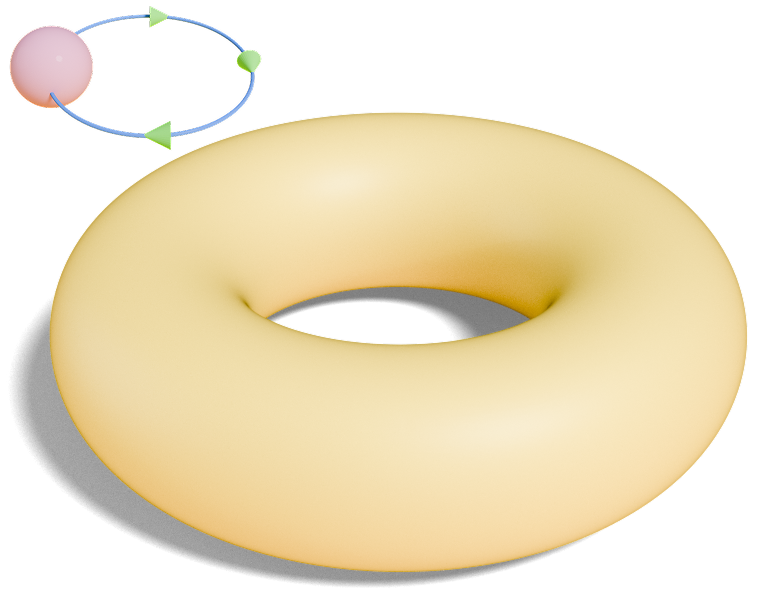}
    \end{overpic}
    \caption{\CG{Swept volume generated by a sphere traversing a circular trajectory. This configuration yields an analytically defined surface, facilitating quantitative evaluation through direct point sampling without explicit construction of a reference model.}}
    \label{fig:analyticalSolution}
\end{figure}

\subsection{Comparisons}
\textbf{Visual Comparison.}
Figure~\ref{fig:compare_expirement} visually compares swept volumes generated by our method against Stamping and \cite{10.1145/3450626.3459780} across various models and motion types.
\CG{Due to varying input parameter formats across different methods, we limit the motion to translation to maintain trajectory consistency.}
The Stamping method exhibits noticeable zigzag artifacts due to discrete temporal sampling. The method of \cite{10.1145/3450626.3459780}, while maintaining temporal continuity via optimization-based distance computations, fails to capture complex geometric features due to its reliance on a single distance field. Our method, employing motion perspective inversion and 4D incremental cutting, preserves geometric details effectively, achieving superior accuracy with reduced computation time.

\noindent
\textbf{Quantitative Comparison.}
For quantitative evaluation, we selected a case with a known analytical solution: a sphere moving along a circular trajectory (Figure~\ref{fig:analyticalSolution}). We sampled points directly from the analytically defined swept surface to measure Chamfer Distance ($L_1$ CD) and Hausdorff Distance (HD). Table~\ref{table:comparison} summarizes these metrics, demonstrating that our method outperforms baseline approaches.

\begin{table}
\caption{Quantitative comparison for different methods.
\textbf{Bold values} indicate optimal performance.
\CG{HD and CD are normalized by the bounding box diagonal length.}
}
\centering
\resizebox{0.7\columnwidth}{!}{%
  \renewcommand{\arraystretch}{1.25}
  \setlength{\tabcolsep}{7pt}
  \rowcolors{2}{white}{rowgray}
  \begin{tabular}{c|ccc}
  \hline
  \rowcolor{headerblue} & Stamping &\cite{10.1145/3450626.3459780} & Ours \\
  \hline
 \CG{CD~(\textperthousand)} $\downarrow$   &       \CG{0.6801}    &        \CG{0.1421}                  &    \CG{\textbf{0.1345}}   \\
  \CG{HD~(\%)} $\downarrow$ &        \CG{1.199}   &         \CG{0.680}                  &    \CG{\textbf{0.645}}   \\
  
  Time (s) $\downarrow$&  181        &             88.3             &   \textbf{80.0}  \\
  \hline
  \end{tabular}%
}
\vspace{-2mm}
\label{table:comparison}
\end{table}
\begin{figure}[htbp]
    \centering
     \begin{overpic}[width=1.0\columnwidth]{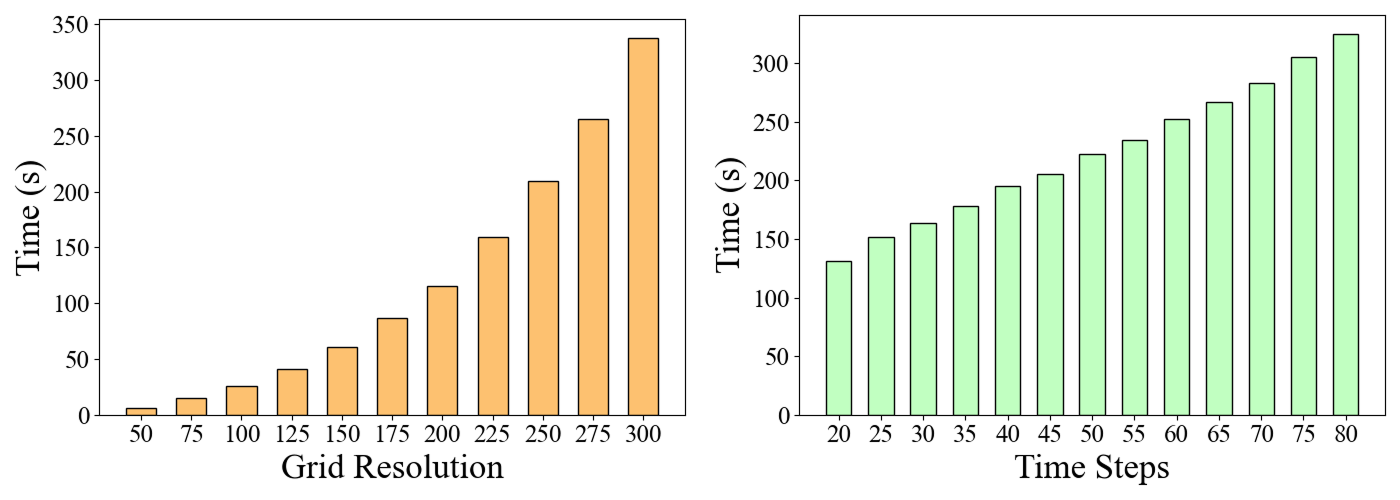}
    \end{overpic}
    \caption{Performance with respect to spatial and temporal discretization. Results represent average computation times across 100 test models.}
    \label{fig:resolution}
\end{figure}
\subsection{Time}
The computational performance of our algorithm is governed by two primary discretization parameters: spatial resolution and temporal sampling density. To quantify their impact on algorithm efficiency, we conducted a comprehensive performance analysis using a dataset of 100 randomly selected geometric models \CG{under simple motions involving translation and rotation}. We measured execution time across systematically varied parameter configurations to characterize performance scaling properties. For spatial resolution experiments, we maintained a fixed temporal discretization of $50$ intervals while varying the spatial grid density. Conversely, for temporal discretization analysis, we established a constant spatial resolution of $256^3$ voxels while adjusting the number of temporal intervals. Figure~\ref{fig:resolution} presents the empirical performance data, illustrating computational complexity as a function of these fundamental discretization parameters.

\begin{figure}[htbp]
    \centering
     \begin{overpic}[width=1.0\columnwidth]{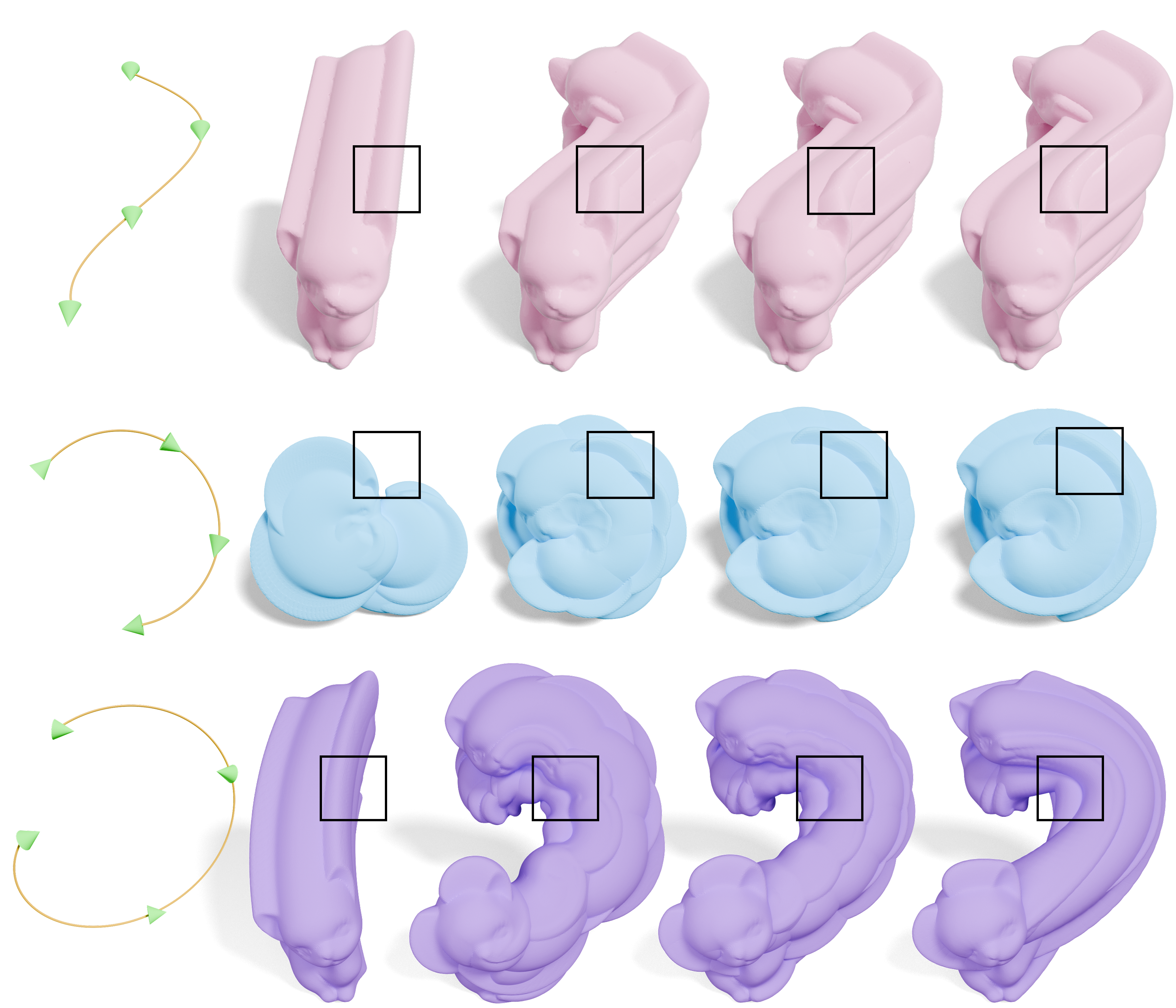}
      \put(25,84){\textbf{\small \CG{$t:26.478s$}}}
      \put(46,84){\textbf{\small \CG{$t:41.701s$}}}
      \put(66,84){\textbf{\small \CG{$t:55.905s$}}}
      \put(85,84){\textbf{\small \CG{$t:78.628s$}}}

      \put(23,51){\textbf{\small \CG{$t:62.592s$}}}
      \put(43,51){\textbf{\small \CG{$t:113.781s$}}}
      \put(63,51){\textbf{\small \CG{$t:142.039s$}}}
      \put(83,51){\textbf{\small \CG{$t:157.657s$}}}
      
      \put(23,30){\textbf{\small \CG{$t:53.415s$}}}
      \put(40,30){\textbf{\small \CG{$t:110.417s$}}}
      \put(61,30){\textbf{\small \CG{$t:119.773s$}}}
      \put(83,30){\textbf{\small \CG{$t:141.930s$}}}

      \put(27,-3){$\mathbf{1}$}
      \put(42,-3){$\mathbf{5}$}
      \put(61,-3){$\mathbf{10}$}
      \put(84,-3){$\mathbf{20}$}

    \end{overpic}
    \caption{\CG{Swept volumes computed at different temporal resolutions with spatial resolution fixed at 256$^3$.}}
    \label{fig:melt_expirement_time}
\end{figure}
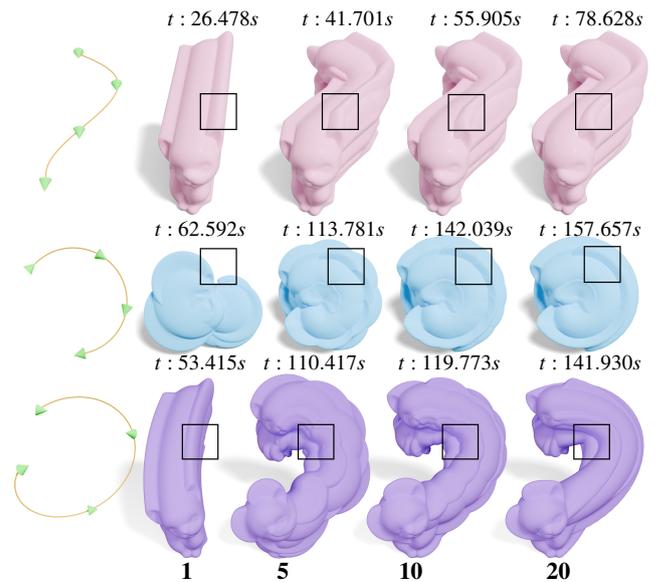
\subsection{Ablation Study}
Our method contains two critical hyperparameters: 1) The temporal discretization parameter, determining the number of time intervals for piecewise linear motion approximation. 2) The spatial resolution parameter, controlling the grid density for extraction.
\CG{Additionally, we include swept volume computation times for all different parameter settings.}

\noindent
\textbf{Time Steps.}
Temporal discretization resolution directly influences the fidelity of our piecewise linear approximation and consequently the accuracy of computed swept volumes.
Figure~\ref{fig:melt_expirement_time} illustrates the resultant swept volumes at temporal resolutions of 1, 5, 10, and 20 discrete intervals. At minimal temporal subdivision, the piecewise linear approximation becomes invalid as an accurate representation of the continuous motion, resulting in substantial geometric deviation from the theoretical surface. Progressive refinement of the temporal discretization demonstrates convergent behavior toward the analytical solution, with notable improvements in geometric continuity at inter-interval boundaries.

\begin{figure}[htbp]
    \centering
     \begin{overpic}[width=1.0\columnwidth]{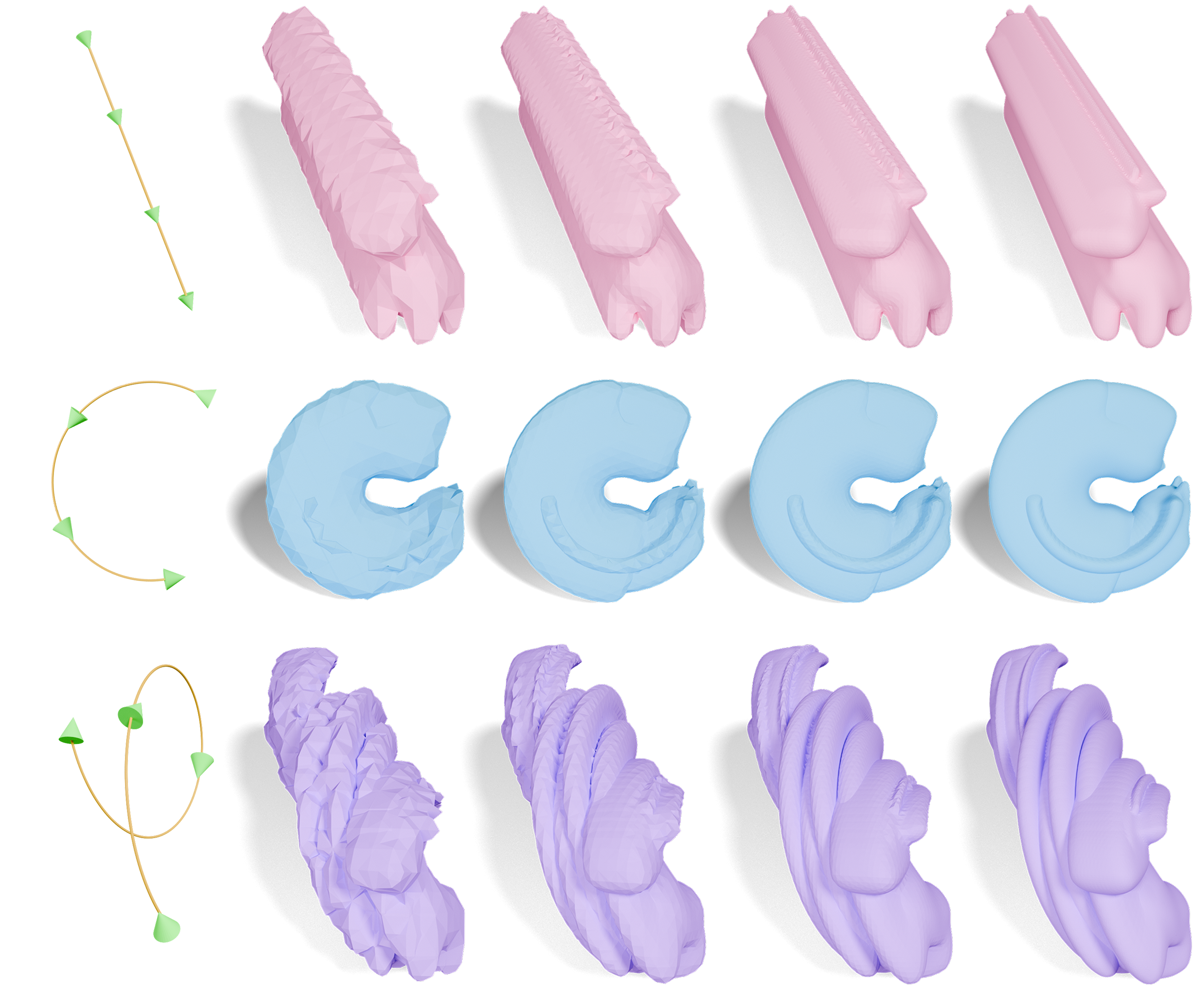}
   
     \put(19,82){\textbf{\small \CG{$t:1.732s$}}}
      \put(38,82){\textbf{\small \CG{$t:6.503s$}}}
      \put(58,82){\textbf{\small \CG{$t:28.392s$}}}
      \put(78,82){\textbf{\small \CG{$t:188.728s$}}}
      
      \put(23,51){\textbf{\small \CG{$t:1.199s$}}}
      \put(44,51){\textbf{\small \CG{$t:4.219s$}}}
      \put(63,51){\textbf{\small \CG{$t:18.795s$}}}
      \put(83,51){\textbf{\small \CG{$t:155.696s$}}}
      
      \put(19,29){\textbf{\small \CG{$t:1.783s$}}}
      \put(39,29){\textbf{\small \CG{$t:5.261s$}}}
      \put(59,29){\textbf{\small \CG{$t:43.106s$}}}
      \put(79,29){\textbf{\small \CG{$t:221.014s$}}}

      \put(31,-4){$\mathbf{25^3}$}
      \put(51,-4){$\mathbf{50^3}$}
      \put(71,-4){$\mathbf{100^3}$}
      \put(91,-4){$\mathbf{256^3}$}
    \end{overpic}
    \vspace{2mm}
    \caption{\CG{Extracted sweep volume at different spatial resolutions with temporal resolution fixed at 50. Flat rendering is used to better highlight the differences across resolutions.}}
    \label{fig:melt_expirement_grid}
\end{figure}

\noindent
\textbf{Grid Resolution.}
Figure~\ref{fig:melt_expirement_grid} presents swept volume extraction results at multiple spatial discretization resolutions:$25^3$, $50^3$, $100^3$, and $256^3$ voxels.
Higher spatial resolution enables more precise approximation of the continuous distance fields, resulting in improved boundary representation and preservation of fine geometric details that would otherwise be lost at coarser discretization levels.

\begin{figure}[htbp]
\centering
\begin{overpic}[width=1.0\columnwidth]{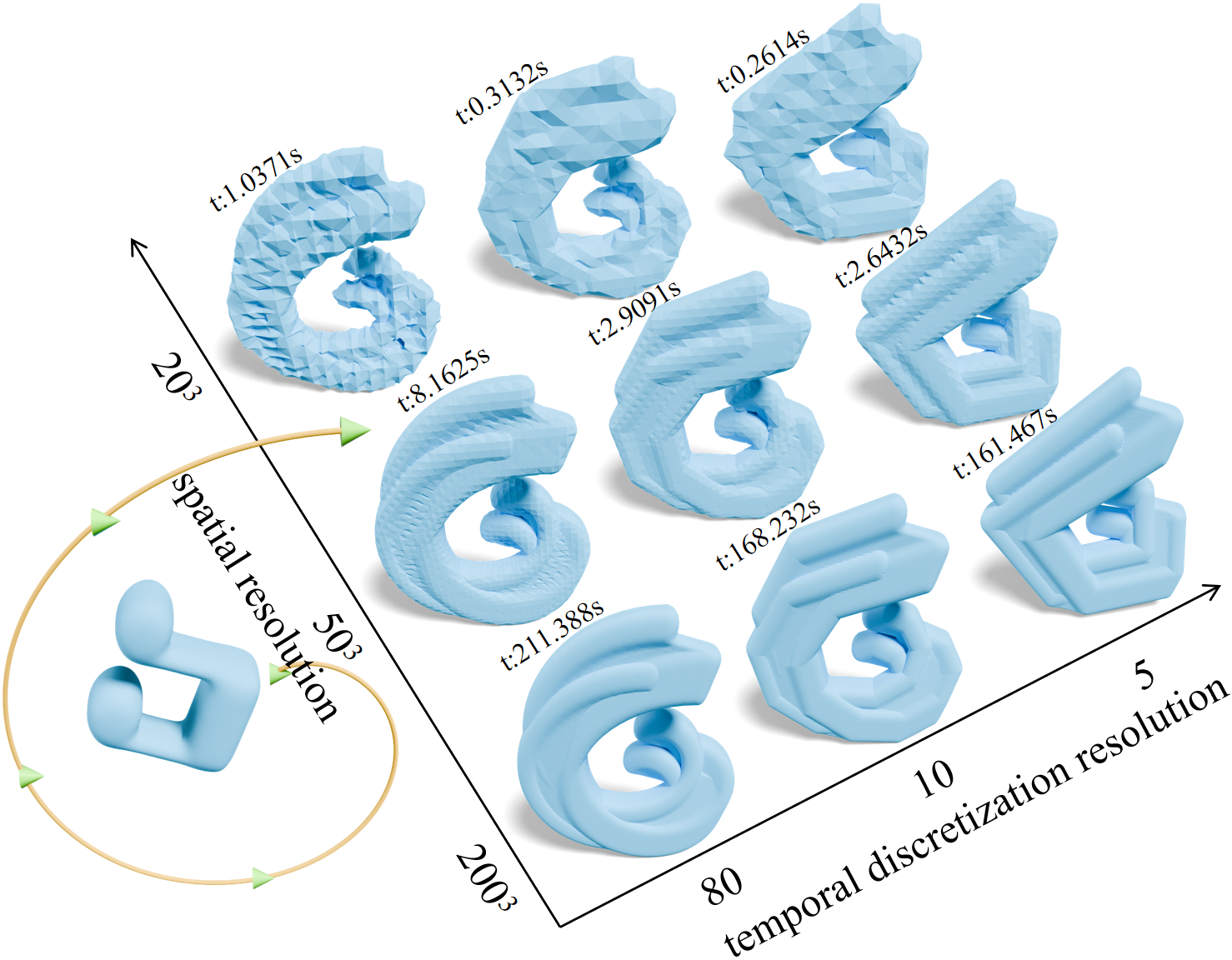}
\end{overpic}
\caption{\CG{Visual comparison across different temporal and spatial discretization resolutions. Flat rendering is used to better highlight the differences across resolutions.}}
\label{fig:spatialResolutionsAndSpatialDiscretization}
\end{figure}

\CG{Figure~\ref{fig:spatialResolutionsAndSpatialDiscretization} demonstrates results in a 3×3 grid where each row and column represents different spatiotemporal resolution settings.}
Additional examples showcasing our method's robustness with various models and motion patterns are provided in Figure~\ref{fig:gallery}.

\section{LIMITATIONS AND FUTURE WORK}

\CG{Our method currently has three primary limitations that represent important directions for future research.}

\noindent
\textbf{Temporal Discretization.} 
\CG{Our method currently employs uniform temporal discretization, dividing the entire motion into equally-sized time intervals. This represents a significant limitation, as it inefficiently allocates computational resources by treating periods of slow, smooth motion the same as rapid or complex motion phases. Adaptive temporal discretization would offer substantial benefits—using fewer time segments during slow movements where linear approximation is highly accurate, and employing finer discretization during rapid motions where more precision is required. Such an approach could significantly improve both computational efficiency and result quality while reducing artifacts at segment boundaries.}

\noindent
\textbf{Spatial Resolution.} \CG{Similarly, our framework relies on uniform spatial tessellation throughout the computational domain. This limitation becomes apparent when considering that swept volume surfaces contain both geometrically complex regions requiring high resolution and relatively flat areas where coarse tessellation would suffice. Adaptive spatial discretization could potentially yield substantial performance gains by concentrating computational resources in complex regions. However, this presents fundamental challenges in our framework since we extract swept volume surfaces independently from each tetrahedron. Adaptive tessellation would cause misalignment issues between adjacent elements with different resolutions, leading to inconsistent surface reconstruction. Investigating hybrid approaches that maintain surface consistency while enabling selective spatial refinement represents an important avenue for future research.}

\noindent
\textbf{Complex Transformations.} \CG{Our inverse trajectory analysis framework assumes that a unified inverse transformation can be applied to spatial points. However, when objects undergo scaling, animation, or deformation, different regions of the model exhibit distinct motion behaviors, making unified inverse transformation impossible. More complex motions such as animation and deformation present fundamental challenges to our current approach. Extending support for such transformations remains an important direction for future work.
}

\section{CONCLUSION}
In this paper, we introduced a novel swept volume computation framework leveraging motion perspective inversion and a multi-field extraction strategy. By fixing the object and analyzing inverse point trajectories, our approach simplifies the complex problem of trajectory analysis. Furthermore, our framework maintains multiple distance fields within each tetrahedral element and utilizes four-dimensional incremental cutting, effectively capturing intricate geometric features emerging during motion. Extensive experiments across a variety of models and motion types validate the superior effectiveness and efficiency of our proposed method compared to existing approaches.

\section*{Acknowledgements}
The authors would like to thank the anonymous reviewers for their valuable comments and suggestions. This work was supported by the National Key R\&D Program of China (2022YFB3303200), and the National Natural Science Foundation of China (U23A20312, 62272277).

\bibliographystyle{eg-alpha-doi} 
\bibliography{main}       


\end{document}